\documentclass[twocolumn,english,aps,prl,superscriptaddress,floats,nobibnotes]{revtex4-1}
\usepackage{amsmath}
\usepackage{amssymb}
\usepackage{amsthm}
\usepackage{setspace}
\usepackage{graphicx}
\usepackage{braket}
\usepackage{mathrsfs}
\usepackage{float}
\usepackage[colorlinks = true,linkcolor = red,urlcolor  = blue,citecolor = blue,anchorcolor = blue]{hyperref}
\usepackage[utf8]{inputenc}
\usepackage[english]{babel}
\usepackage{bm}

\begin{document}

\title{Anomalous transport in Kane fermions}

\author{Karun Gadge}
\affiliation{School of Basic Sciences, Indian Institute of Technology Mandi, Mandi 175005, India}
\author{Sumanta Tewari}
\affiliation{Department of Physics and Astronomy, Clemson University, Clemson, South Carolina 29634, USA}
\author{Girish Sharma}
\affiliation{School of Basic Sciences, Indian Institute of Technology Mandi, Mandi 175005, India}

\begin{abstract}
Kane fermions are characterized by a linear Dirac cone intersecting with a flat band, resembling a pseudo-spin-1 Dirac semimetal.
Similar to relativistic Dirac fermions, Kane fermions satisfy a linear energy-momentum relation and can be classified as being pseudo-relativistic. Though not protected by symmetry or by topology, Kane fermions can emerge by suitable band engineering, for example, in mercury-telluride compounds. Here we study the Berry curvature of Kane fermions that emerges in the presence of time-reversal symmetry breaking weak Zeeman fields. We  discuss the related anomalous transport coefficients and discuss the anisotropy in these responses that can be probed in experiments. 
\end{abstract}

\maketitle

\textit{Introduction:}
A linear energy-momentum relation has been of much interest to condensed matter physicists as it mimics the relativistic behavior of Dirac particles in high-energy physics~\cite{peskin1995introduction}. 
Topological three-dimensional  Dirac semimetals (DSM)~\cite{murakami2007phase,murakami2007tuning,young2012dirac,yang2014classification} and three-dimensional Kane fermions~\cite{orlita2014observation} are prominent examples of materials where signatures of a linear energy-momentum relation have been observed. DSM   is a  stable 3D electron system with fourfold degenerate bulk Dirac nodes protected by crystalline symmetries that can arise at the quantum critical point between a 3D topological insulator and a conventional insulator with fine tuning of an external parameter~\cite{murakami2007phase,murakami2007tuning,young2012dirac,yang2014classification}. DSMs are closely related to topological Weyl semimetals (WSM) that are three-dimensional materials, where the twofold degenerate Weyl nodes in the bulk
energy spectrum are stable due to the existence of a nonzero
Chern number invariant associated with each Weyl node~\cite{wan2011topological,burkov2011weyl,xu2011chern}. In DSMs, time-reversal and space-inversion symmetry are simultaneously preserved which ensures that the Chern number vanishes for each Dirac node, while WSMs  violate at least one of the symmetries. Weyl nodes with threefold as well as fourfold degeneracies have been also predicted~\cite{bradlyn2016beyond} with the low-energy Hamiltonian  of the form $\mathbf{k}\cdot \mathbf{S}$, where $\mathbf{S}$ is the vector of spin-1 or-3/2 matrices. Such nodes are protected by higher Chern numbers. 
A substantial amount of current experimental effort is being directed to the study of novel topological features in these materials such as quantum Hall effect, anomalous Hall and Nernst effects, planar Hall effect, chiral anomaly among several others~\cite{yang2011quantum,burkov2014anomalous,sharma2016nernst,sharma2017nernst,liang2017anomalous,li2016chiral,wang2018large,adler1969axial,nielsen1981no,nielsen1983adler,bell1969pcac,aji2012adler,zyuzin2012weyl,zyuzin2012weyl,son2012berry,goswami2013axionic,goswami2015optical, fukushima2008chiral, lundgren2014thermoelectric}. 

A distinct class of fermionic quasiparticles, namely Kane fermions, has been recently discovered in Hg$_{1-x}$Cd$_x$Te crystals existing at the critical cadmium concentration of  $x_c\sim 0.17$~\cite{orlita2014observation}. These emerge at the boundary of a topological phase transition from a semiconductor ($x>x_c$) to a semimetal ($x<x_c$). Kane fermions are characterized by a Dirac cone intersected by an additional flat band, closely resembling a spin-1 Dirac semimetal~\cite{bradlyn2016beyond}. However, unlike Dirac and Weyl fermions, Kane fermions are not protected by symmetry or topology, but rather the band-structure of these compounds can be suitably engineered according to interest. The three-dimensional pseudo-relativistic gas of massless Kane fermions has been confirmed in magneto-optical measurements~\cite{orlita2014observation} and temperature dependent far-infrared magneto-spectroscopic studies~\cite{Teppe2016} on Hg$_{1-x}$Cd$_x$Te, while its magneto-optical signatures have been observed in Cd$_3$As$_2$~\cite{akrap2016magneto}. 

On the level of band-structure, Kane fermions and a  spin-1 Dirac semimetal are identical, i.e., doubly degenerate bands with dispersion $E=\pm\hbar v_F k$ intersect with a degenerate flat band of exactly zero energy. The topology of the bands is however quite different in the two cases.  
It is  possible to express the low-energy effective Hamiltonian of a DSM as two copies of WSM with opposite Chern numbers, but such a decomposition is not typically possible for Kane fermions. Nevertheless, Kane fermions have some hidden topological features that is demonstrated by the fact that the two-dimensional sectors of Kane fermions (in the limit $k_z=0$) has been linked to the $\alpha-T_3$ model~\cite{malcolm2015magneto,raoux2014dia} exhibiting non-quantized Berry phase. Thus on phenomenological grounds, it possible to explore the hidden topological nature of Kane fermions by breaking either time-reversal or spatial inversion symmetry that will generate a non-vanishing flux of Berry curvature. 

In this Letter, we study the band-structure and the generated Berry curvature in the six-band Kane fermion model in the presence of weak Zeeman fields (such that Landau quantization is unimportant) that break time-reversal symmetry. We find that similar to the case of Weyl fermions, the Berry curvature can diverge at certain points in the Brillouin zone giving rise to anomalous transport responses. On the other hand, the distribution of the flux is distinct from the case of a WSM. We also study semiclassical transport of Kane fermions and focus on the anomalous Hall and anomalous Nernst effects that can be experimentally probed in the limit of weak Zeeman fields. We find anisotropy in responses depending on the direction of the field. We also present a toy model of a relativistic three band system that exhibits anisotropic responses despite its isotropic relativistic band dispersion.   

\begin{figure}
    \centering
    \includegraphics[width=1\columnwidth]{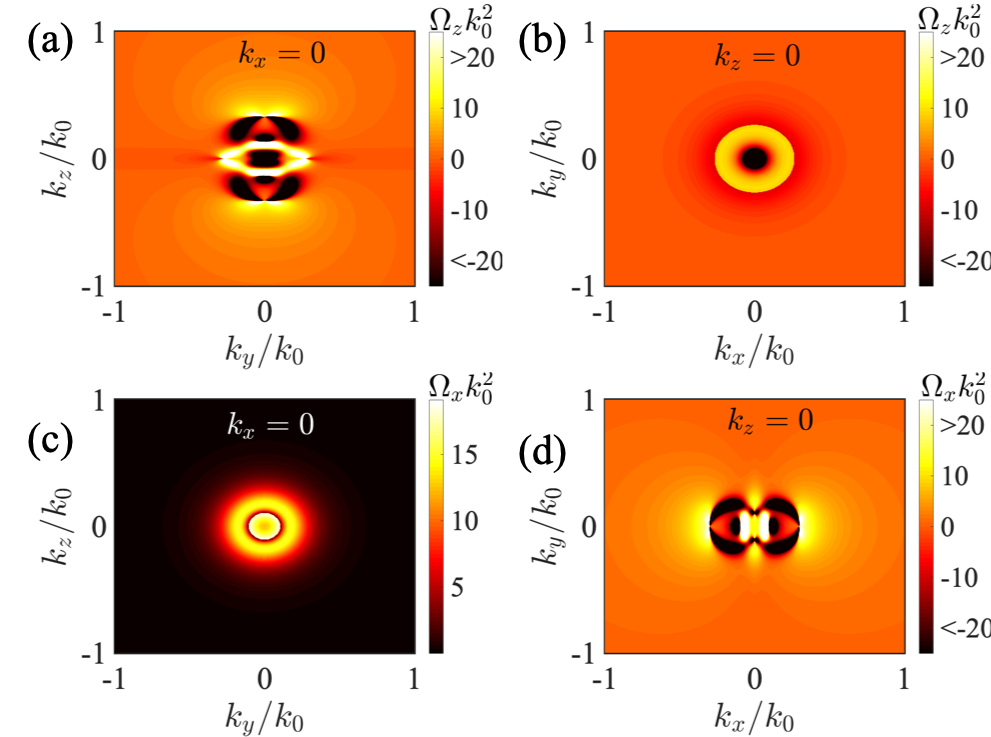}
    \caption{Components of the Berry curvature for the  lowest energy  band plotted along various planes when the field is applied along the $z-$axis (a) and (b) and applied along the $x-$axis (c) and (d).}
    \label{fig:berry_1}
\end{figure}

\textit{Six-band Kane fermion model:} Up to first-order in $\mathbf{k}\cdot\mathbf{p}$ theory, the effective Hamiltonian in the basis $\psi_{\mathbf{k}}=(c_{\mathbf{k},\uparrow},c_{\mathbf{k},3/2},c_{\mathbf{k},-1/2},c_{\mathbf{k},\downarrow},c_{\mathbf{k},-3/2},c_{\mathbf{k},1/2})$ is given by~\cite{orlita2014observation} 
\begin{align}
&\mathcal{H}_\mathbf{k} =  \begin{pmatrix}
0 & \frac{\sqrt{3}vk_+}{2} & -\frac{vk_-}{2}&0&0&-vk_z\\
\frac{\sqrt{3}vk_-}{2} & 0 &0&0&0&0\\
-\frac{vk_+}{2} &0 &0&-vk_z&0&0\\
0&0&-vk_z&0&-\frac{\sqrt{3}vk_-}{2}&\frac{vk_+}{2} \\
0&0&0&-\frac{\sqrt{3}vk_+}{2}&0&0\\
-vk_z&0&0&\frac{vk_-}{2}&0&0\\
\end{pmatrix}
\label{Eq_Hk1}
\end{align} 
The operators $c_{\mathbf{k},\uparrow/\downarrow}$ destroy spin-up/down electrons in the conduction band, while the other operators destroy electrons in the valence band of corresponding $z$-component of the total angular momentum. In the above Hamiltonian $v$ is the Fermi velocity, and $k_{+/-}=k_x\pm i k_y$. One can readily verify that the Hamiltonian has three doubly-degenerate eigenvalues: 0,$\pm k$. The Hamiltonian respects time-reversal symmetry $\sigma_y \mathcal{H}^*_\mathbf{k}\sigma_y=\mathcal{H}_{-\mathbf{k}}$.
For broken time-reversal symmetry, the band-structure is in general complicated and depends on the direction of magnetic field as well as on the $g-$factors of the problem. To this end we use an effective model~\cite{cano2017chiral} for the Zeeman coupling where we add to the Hamiltonian in Eq.~\ref{Eq_Hk1} the Zeeman term 
$\mathcal{H}_z=\mathbf{b}\cdot \mathbf{J}$, where $\mathbf{b}$ is the Zeeman field, $J_x$, $J_y$, $J_z$ are the following matrices: $J_z=m_z\otimes\sigma_z$, $J_x=m_x\otimes\sigma_x$, and $J_y=m_y\otimes\sigma_y$, where $\sigma_i$'s are Pauli matrices, and
\begin{align}
m_z&=\left(
\begin{array}{ccc}
 1 & 0 & 0 \\
 0 & \frac{3}{2} & 0 \\
 0 & 0 & -1 \\
\end{array}
\right);m_x=\left(
\begin{array}{ccc}
 1 & 0 & 0 \\
 0 & 0 & \frac{\sqrt{3}}{2} \\
 0 & \frac{\sqrt{3}}{2} & 1 \\
\end{array}
\right);\nonumber\\ m_y&=\left(
\begin{array}{ccc}
 -1 & 0 & 0 \\
 0 & 0 & \frac{\sqrt{3}}{2} \\
 0 & \frac{\sqrt{3}}{2} & -1 \\
\end{array}
\right)
\end{align}
\begin{figure}
    \centering
    \includegraphics[width=1\columnwidth]{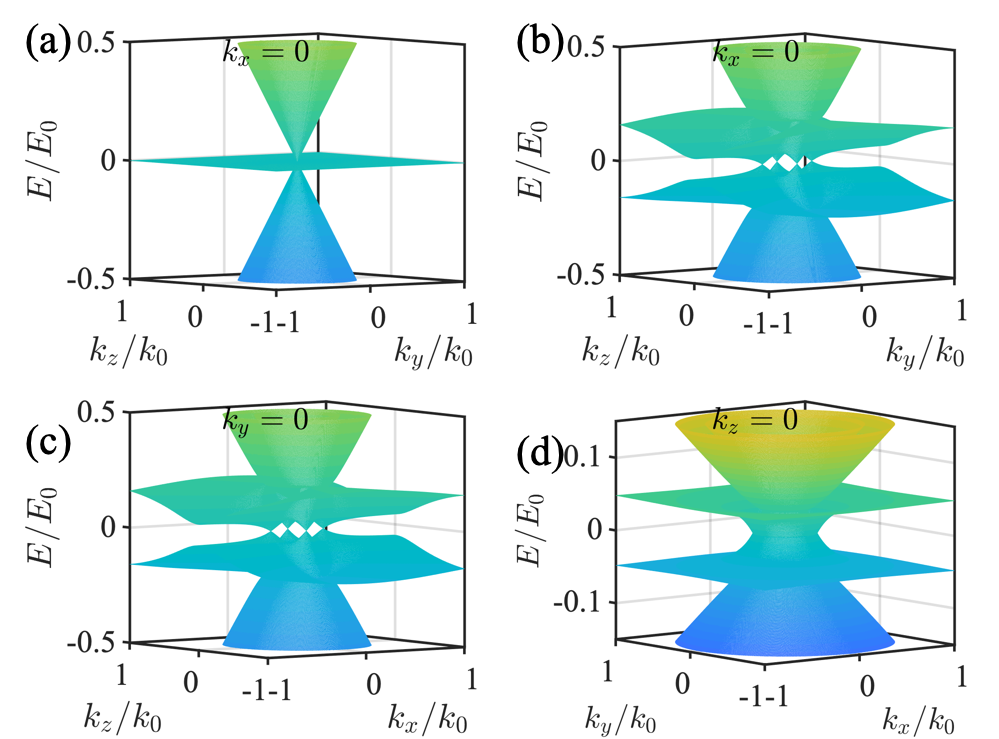}
    \caption{(a) Band-structure of  Kane fermions in the absence of Zeeman field. (b)-(d) Band-structure when Zeeman field is applied along the $z-$axis shown along various planes.}
    \label{fig:bs1}
\end{figure}
In the absence of Zeeman field the bandstructure consists of three doubly-degenerate bands $E_\mathbf{k}=0,\pm k$, as shown in Fig.~\ref{fig:bs1}(a). In the presence of Zeeman field the band-degeneracy is broken as also shown in Fig.~\ref{fig:bs1}(b)-(d). On inspection we find that the low-energy bands appear to touch each other at multiple points in the Brillouin zone. Since closed form expressions for the energy bands in the presence of Zeeman field do not exist, therefore we resort to numerical evaluation of the band structure. The detailed description of how the bandstructure  evolves when the  Zeeman field is applied across various directions is presented in Ref.~\cite{Supplemental}.
Zeeman field breaks time-reversal symmetry and generates a flux of Berry curvature given by 
\begin{align}
    \Omega^{n}_{\gamma} =i \sum_{n' \neq n} \frac{\langle n |d\mathcal{H}/dk^\alpha| n' \rangle \langle n' |d\mathcal{H}/dk^\beta|  n \rangle - (\alpha\leftrightarrow\beta)}{(E^n_\mathbf{k} - E^{n'}_\mathbf{k})^2},
\end{align}
which becomes giant when the bands almost touch each other as shown in Fig.~\ref{fig:berry_1}. 
Simultaneously, the self-rotation of the Bloch wave-packet also generates an intrinsic orbital magnetic moment given by~\cite{xiao2010berry} 
\begin{align}
        m_{\mathbf{k}}&=\frac{-i e}{2\hbar}\bra{\nabla_{\mathbf{k}} n }\mathbf{\times}[\mathcal{H}\mathbf{(k)}-E_\mathbf{k}]\ket{\nabla_\mathbf{k} n}
\end{align}
The anomalous magnetic moment couples to the applied magnetic field as $E_\mathbf{k}\rightarrow E_\mathbf{k} - \mathbf{m}_\mathbf{k}\cdot\mathbf{B}$.

\textit{Transport responses:} Here we focus on the anomalous transport responses that are specifically generated by the presence of Berry curvature due to broken time-reversal symmetry. Specifically, we calculate the following  responses: (i) anomalous Hall effect $\sigma_{xy}$, and (ii) the anomalous Nernst effect. All of these effects can be measured experimentally and have been studied for Weyl and Dirac semimetals~\cite{li2016chiral,liang2017anomalous,wang2018large}. This provides a strong motivation to experimentally look for these effects in Kane fermionic systems as well. Even though Kane fermions are topologically distinct from Weyl or Dirac fermions, the anomalous effects must prevail on phenomenological grounds due to the presence of a finite flux of Berry curvature that  diverges at certain points in the Brillouin zone. 

The presence of a giant flux of the Berry curvature due to broken time-reversal symmetry yields a purely anomalous Hall effect that is given by
\begin{align}\sigma_{\alpha\beta}=\epsilon_{\alpha\beta\gamma}\sum\limits_n\frac{e^2}{\hbar}\int_{}^{}\frac{d^3k}{(2\pi)^3}\Omega_\gamma^n{f}_{\mathrm{eq}}(\mathbf{k}),
\end{align}
where $f_{\mathrm{eq}}(\mathbf{k})$ is the Fermi-Dirac equilibrium distribution function. 
\begin{figure}
    \centering
    \includegraphics[width=\columnwidth]{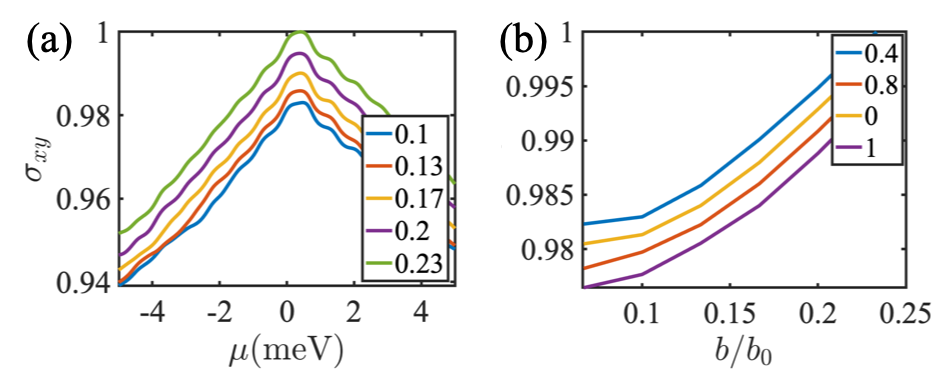}
    \caption{Normalized anomalous Hall conductivity $\sigma_{xy}$ as (a) a function of the chemical potential $\mu$ for various values of the Zeeman field ($b/b_0$) applied along the $z-$axis, and (b) function of the Zeeman field at different chemical potentials (in meV).}
    \label{fig:sxy1}
\end{figure}
Fig.~\ref{fig:sxy1} plots this as a function of the chemical potential as well as the Zeeman field. The anomalous Hall conductivity peaks near $\mu=0$ as in the case of Weyl semimetals~\cite{goswami2013axionic,sharma2016nernst} due to the Berry curvature becoming singular at the band touching points. The slight offset of the peak from $\mu=0$ is due to the energy shift acquired due to the orbital magnetic moment contribution. The behavior of the anomalous Hall effect with the chemical potential is similar to that observed in Weyl semimetals~\cite{goswami2013axionic}. As a side note we point out that the small  oscillations in $\sigma_{xy}$ are more of a numerical artifact as their magnitude decreases on reducing the integration bin size as demonstrated in Ref.~\cite{Supplemental}.
The Hall conductivity also increases with the Zeeman field for relevant ranges of the field. 

Another related phenomena is the anomalous Nernst effect. In linearized models of Weyl semimetals, while the anomalous Hall effect survives, the anomalous Nernst effect is predicted to vanish~\cite{lundgren2014thermoelectric}, though it still survives in more realistic models~\cite{sharma2016nernst,sharma2017nernst}. The anomalous Nernst coefficient is calculated as 
\begin{align}
\alpha_{\alpha\beta}=\epsilon_{\alpha\beta\gamma}\sum\limits_n\frac{e^2}{\hbar}\int_{}^{}\frac{d^3k}{(2\pi)^3}\Omega_\gamma^n{s}_\mathbf{k},
\end{align}
where $s_{\mathrm{\mathbf{k}}} = -f_{\mathrm{eq}}(\mathbf{k}) \log (f_{\mathrm{eq}}(\mathbf{k}) ) - (1-f_{\mathrm{eq}}(\mathbf{k}) )\log (1-f_{\mathrm{eq}}(\mathbf{k}) )$ is the entropy density of the Kane gas. 
\begin{figure}
    \centering
    \includegraphics[width=\columnwidth]{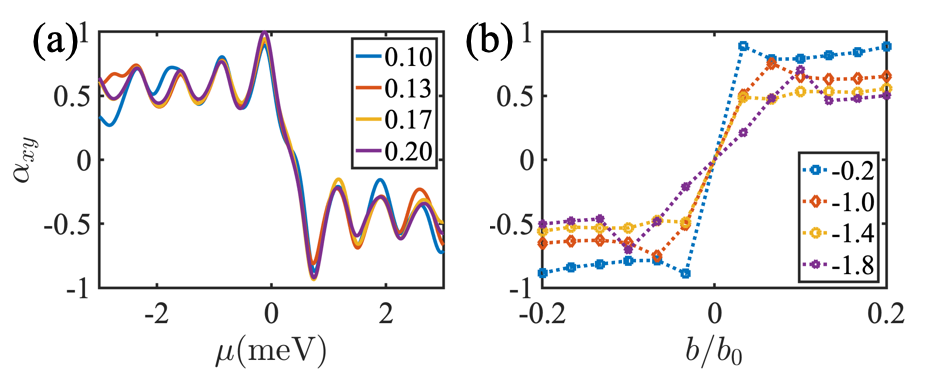}
    \caption{Normalized anomalous Nernst conductivity $\alpha_{xy}$ as (a) a function of the chemical potential $\mu$ for various values of the Zeeman field ($b/b_0$) applied along the $z-$axis, and (b) function of the Zeeman field at different chemical potentials (in meV).}
    \label{fig:axy}
\end{figure}
Fig.~\ref{fig:axy} plots the anomalous Nernst coefficient $\alpha_{xy}$ as a function of the chemical potential $\mu$ for various values of the Zeeman field. The anomalous Nernst coefficient peaks slightly above and below the nodal point (at $\mu=0$) and is not much sensitive to the magnitude of the Zeeman field (because it is related to the derivative of $\sigma_{xy}$. Note that for the same reason the tiny fluctuations in $\sigma_{xy}$ are seen to be magnified in $\alpha_{xy}$). The behavior of $\alpha_{xy}$ with respect to the Zeeman field shows a step-like profile near $b=0$ as also observed in the Nernst coefficient in Dirac semimetals~\cite{liang2017anomalous,sharma2017nernst}.
\begin{figure}
    \centering
    \includegraphics[width=\columnwidth]{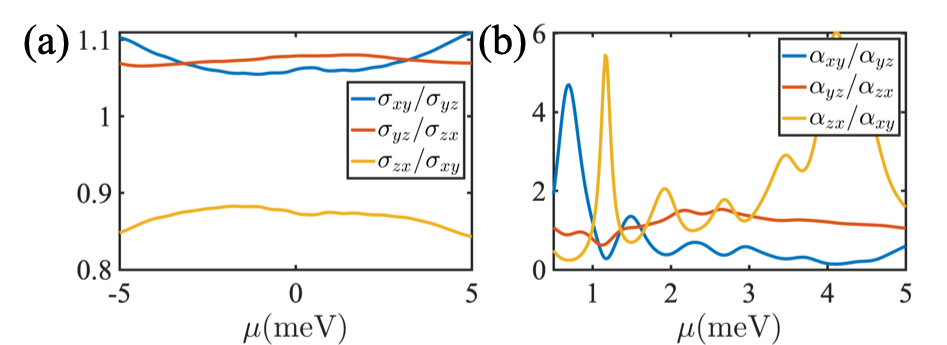}
    \caption{Anisotropy in the anomalous Hall conductivity (a) and the anomalous Nernst conductivity (b) when the Zeeman field applied across various directions (if the field is applied across the $z-$axis, the response is measured in the plane perpendicular to it, i.e. $\sigma_{xy}$) and so on. We chose $b/b_0=0.07$. }
    \label{Fig:anisotropy}
\end{figure}

Fig.~\ref{Fig:anisotropy} compares the anisotropy in the anomalous Hall and Nernst conductivities when the Zeeman field is applied across various directions. For example, if the field is applied across the $z-$axis, the response is measured in the plane perpendicular to it (i.e. $\sigma_{xy}$) and so on. We note that even though the energy dispersion is isotropic along all the three directions, both the anomalous Hall conductivity and the anomalous Nernst conductivity can show significant anisotropic response depending on the direction of the applied field. This is contrasted to the case of a spin-1/2 or a spin-1 Weyl fermion  that generates purely isotropic response. The anisotropy arises from the nature of the wavefunctions and the geometrical Berry phase that is no longer identical across the three axes. Experiments on these systems will clearly be able to measure the anomalous responses and detect the anisotropic nature of the Kane gas under broken time-reversal symmetry. 
As a matter of principle, we show below how anisotropic responses can  arise from relativistic  isotropic dispersions. 

\textit{Relativistic fermions with hidden anisotropy:}  The generic $\mathbf{k}-$space Hamiltonian for a chiral Weyl fermion, hosting a $2n+1$ fold degenerate Weyl point is given by
\begin{align}
H_\mathbf{k} =\mathbf{k}\cdot  \mathbf{S},
\label{Eqn_Hk_1}
\end{align}
where $\mathbf{S}$ is a vector of $2n+1$ dimensional matrices, satisfying the algebra $[S_\alpha, S_\beta] = i\epsilon_{\alpha\beta\gamma} S_\gamma$, and $n\in \{\mathbb{Z}/2 \}$. We can choose $\mathbf{S}$ to be identical to the spin-$n$ matrices. The energy dispersion, Berry curvature, orbital magnetic moment and all transport responses are isotropic in this case.
Next we consider a prototype model for a 3-band fermionic system, given by 
\begin{align}
H_\mathbf{k} = \mathbf{k}\cdot\boldsymbol{\mathcal{M}}
\label{Eq_Hk_2}
\end{align}
where the matrices $\boldsymbol{\mathcal{M}}$ are the following 
\begin{align}
&\mathcal{M}_x =  \begin{pmatrix}
0 & 1 & 0\\
1 & 0 &0\\
0 &0 &0
\end{pmatrix};
\mathcal{M}_y =  \begin{pmatrix}
0 & \sqrt{3}i/2 & -i/2\\
-\sqrt{3}i/2 & 0 &0\\
i/2 &0 &0
\end{pmatrix};\nonumber\\
&\mathcal{M}_z =  \begin{pmatrix}
0 & 0 & 1\\
0 & 0 &0\\
1 &0 &0
\end{pmatrix}.
\end{align} 
It is easily verified that the above matrices do not satisfy the usual angular momentum algebra, i.e.,  $[\mathcal{M}_\alpha, \mathcal{M}_\beta] \neq i \epsilon_{\alpha\beta\gamma} \mathcal{M}_\gamma$, unlike  the $\mathbf{S}$ matrices in Eq.~\ref{Eqn_Hk_1}. 
Interestingly, the Hamiltonian produces non-degenerate energy bands with relativistic dispersion $E_\mathbf{k} = 0, +k, -k$, which is also the case of relativistic chiral fermions in Eq.~\ref{Eqn_Hk_1}. It is therefore of interest to compare and contrast the properties of the above model with the properties of chiral relativistic fermions.
We first examine the Berry curvature of the above model, which is evaluated to be 
\begin{align}
\Omega_i^{(\pm)} = \frac{{k_i} \left({k_x}+\sqrt{3} {k_z}\right)}{2
	k^4}, \nonumber\\ 
\Omega_i^{(0)} = -\frac{{k_i} \left({k_x}+\sqrt{3} {k_z}\right)}{
	k^4},
\end{align}
where the subscript $i$ is the spatial coordinate and the superscript indicates the band index. It is of interest to note that the Berry curvature of the electron ($E_\mathbf{k} = +k$) and hole ($E_\mathbf{k} = -k$) bands is of the same sign. The net Berry curvature vanishes (as expected) because the flat-band compensates for the Berry curvature (i.e., $\Omega_i^{(0)} = -2 \Omega_i^{(\pm)}$). Clearly  the Berry curvature of the bands is purely anisotropic, in contrast to that of chiral Weyl fermion. A single $n=1$ Weyl node is protected by a Chern number of $\mathcal{C} = \pm 1$, also highlighting the fact that Weyl nodes act as source and sink of the Berry curvature (which diverges at the nodal point). In contrast, the Chern number of the bands of the Hamiltonian in Eq.~\ref{Eq_Hk_2} is evaluated to be exactly zero. Even though the Berry curvature diverges at the nodal point, the nodal point acts both as a source and a sink of the Berry curvature. Thus the nodal point is not topologically protected in this case.

Focusing on the contribution from the electron band (the flat band and the hole band can be evaluated similarly), we find that at $T\rightarrow 0$
\begin{align}
\sigma_{xy} &=\frac{e^2}{\hbar} \frac{1}{(2\pi)^2} \frac{k_F}{\sqrt{3}}\nonumber \\
\sigma_{yz} &=\frac{e^2}{\hbar} \frac{1}{(2\pi)^2} \frac{k_F}{{3}}\nonumber \\
\sigma_{zx} &=0
\end{align}
We note from the above that the anomalous Hall effect is anisotropic in all the three directions, and the anomalous Hall effect is non-zero in two of the directions. This can be again contrasted to the case of a chiral Weyl fermion which shows $\sigma_{\alpha\beta} = 0$ in all the three directions. Note that the anomalous Hall effect in a time-reversal symmetry broken WSM arises from the flux of the Berry curvature flowing from one node to the other and not from a single Weyl node alone.

\textit{Conclusions:}  Weak time-reversal breaking fields can generate a giant flux of Berry curvature in Kane fermions and can give rise to anisotropic anomalous responses such as the anomalous Hall and Nernst effects. It is of utmost experimental interest to probe these responses in Kane fermionic systems. 

\textit{Acknowledgement:} G. S. acknowledges support from SERB Grant No.
IITM/SERB/GS/305. S. T. acknowledges support from Grant No. NSF 2014157.

\bibliographystyle{apsrev4-1}
\bibliography{main.bbl}
\end{document}